\documentstyle{article}

\newfont{\SETT}{msbm10 scaled \magstep4}
\newfont{\SET}{msbm10 scaled \magstep3}
\newfont{\Set}{msbm10 scaled \magstep2}
\newfont{\Settoc}{msbm10 scaled \magstep1}
\newfont{\set}{msbm10}

\newcommand{\beq}{\begin{equation}}
\newcommand{\eeq}{\end{equation}}
\newcommand{\bea}{\begin{eqnarray}}
\newcommand{\eea}{\end{eqnarray}}

\newcommand{\cd}{\partial}

\newcommand{\ra}{\rightarrow}

\newcommand{\R}{\mbox{\set R}}
\newcommand{\Z}{\mbox{\set Z}}

\newcommand{\N}{\mbox{\set N}}

\newcommand{\T}{TD$\phi^{4}$S\ }

\newtheorem{prop}{Proposition}

\begin{document}
\title{A discrete $\phi^{4}$ system without Peierls-Nabarro barrier}
\author{J.M. Speight \\
Department of Mathematics \\
University of Texas at Austin \\
Austin, Texas 78712, U.S.A.}
\date{}
\maketitle

\begin{abstract}
A discrete $\phi^{4}$ system is proposed which preserves the topological
lower bound on the kink energy. Existence of static kink solutions saturating
this lower bound and occupying any position relative to the lattice is
proved. Consequently, kinks of the model experience no Peierls-Nabarro
barrier, and can move freely through the lattice without being pinned. 
Numerical simulations reveal that kink dynamics in this system is significantly
less dissipative than that of the conventional discrete $\phi^{4}$ system, so
that even on extremely coarse lattices 
the kink behaves much like its continuum counterpart. It is argued,
therefore, that this is a natural discretization for the purpose of
numerically studying soliton dynamics in the continuum $\phi^{4}$ model.

\end{abstract}

\section{Introduction}
\label{sec:int}

Solitons have diverse applications in many branches of physics. Those with 
applications in high energy physics are topological solitons which arise
almost exclusively in non-integrable Lagrangian field theories (integrability
appears incompatible with Lorentz invariance in spacetimes of realistic
dimension). In the absence of integrability, there is no hope of solving the
multisoliton initial value problem exactly. One approach to the study of
soliton dynamics is to perform numerical simulations of the field equations.
In so doing one is forced to discretize space in some way, and this inevitably
introduces fictitious discretization effects into the dynamics, which one 
should seek to minimize.

The standard discretization of  a field theory is obtained by replacing
spatial partial derivatives by simple difference operators, leaving it 
otherwise unchanged, so that one is really studying a large network of
oscillators, each moving in an identical substrate potential, with nearest
neighbour couplings. Such systems are intrinsically interesting (they have
applications in condensed matter and biophysics) and have been extensively
studied in recent years \cite{Pey,Boe,Dun}. One discretization effect is that
static solitons can no longer be centred on any point in space: they must
lie in the centre of a lattice cell, or exactly on a lattice site, and the
energies of these two types of static solution are different. So there is an
energy barrier, called the Peierls-Nabarro (PN) potential resisting the
propagation of a soliton from cell to cell. As a soliton moves through the
lattice, its periodic motion in and out of the PN potential excites small
amplitude travelling waves (radiation) which  travel in its wake, draining its
kinetic energy. The soliton decelerates under this dissipation until it no
longer has sufficient energy to surmount the PN barrier. It is then pinned
to a lattice cell.

If one is to use the standard discretization to simulate continuum soliton
dynamics, these extraneous discretization effects must be rendered 
insignificant by using a very fine spatial lattice, with spacing $h$ of order
(soliton width)/20. This works well, but is computationally expensive,
especially in high dimensions. In this paper we consider an alternative
approach, originated by Richard Ward, in the case of $\phi^{4}$ theory in
$1+1$ dimensions. Exploiting the non-uniqueness of the discretization process,
one can find a discrete $\phi^{4}$ system which has no PN barrier. One would
expect kink dynamics in such a system to be much simpler and more
continuum-like than in the conventional discrete system, even on relatively
coarse lattices. This expectation has been confirmed for a similar discrete
sine-Gordon system \cite{Spe}. The rest of the paper is organized as follows:
in section \ref{sec:top} we define the discrete $\phi^{4}$ system, in section
\ref{sec:kin} we prove that it has no PN barrier, in section \ref{sec:num}
we present numerical simulations of the system, and compare its performance 
with that of the conventional discrete system, and finally section 
\ref{sec:con} presents some concluding remarks.

\section{A topological discrete $\phi^{4}$ system}
\label{sec:top}

The continuum $\phi^{4}$ model consists of a real valued scalar field
$\phi:\R^{1+1}\ra\R$ whose time evolution is governed by the Lagrangian
$L=E_{K}-E_{P}$, the kinetic and potential energy functionals being,
respectively,
\bea
E_{K}&=&\int_{-\infty}^{\infty}dx\, \frac{1}{2}\dot{\phi}^{2} \\
E_{P}&=&\int_{-\infty}^{\infty}dx\, 
\left[\frac{1}{2}\left(\frac{\cd\phi}{\cd x}
\right)^{2}+\frac{1}{8}(1-\phi^{2})^{2}\right].
\eea
To ensure that $\phi$ has finite energy, one imposes that $\phi\ra\pm 1$ as
$|x|\ra\infty$. The trivial solutions $\phi(x,t)=\pm 1$ are the vacua of the
model, while a solution interpolating between $-1$ and $1$ ($1$ and $-1$) is
a kink (antikink). Static kinks are minimals of $E_{P}[\phi]$ and can be found
by means of a Bogomol'nyi argument \cite{Bog}. Namely, 
if $\lim_{x\ra\pm\infty}\phi=\pm 1$, then
\bea
0&\leq&\frac{1}{2}\int_{-\infty}^{\infty}dx\, \left[\frac{d\phi}{dx}-
\frac{1}{2}(1-\phi^{2})\right]^{2} 
= E_{P}-\frac{1}{2}\int_{-\infty}^{\infty}dx\, \frac{d\phi}{dx}(1-\phi^{2})
\nonumber \\
&=& E_{P}-\frac{1}{2}\left[\phi-\frac{1}{3}\phi^{3}\right]^{1}_{-1}
=E_{P}-\frac{2}{3}.
\eea
So $E_{P}\geq\frac{2}{3}$ in the kink sector, with equality if and only if
\beq
\label{4}
\frac{d\phi}{dx}=\frac{1}{2}(1-\phi^{2}).
\eeq
Equation (\ref{4}) is called the Bogomol'nyi equation. Note that it is a
first order o.d.e., unlike the static field equation which is second order.
It is easily solved, yielding
\beq
\label{5}
\phi(x)=\tanh\frac{1}{2}(x-b)
\eeq
where the constant of integration $b$ is identified as the position of the
kink centre.

Upon discretization, $x$ takes values in $h\Z=\{0,\pm h,\pm 2h,\ldots\}$,
where $h$ is the lattice spacing. It is convenient to introduce the notation
$f_{+}(x)=f(x+h)$ and $f_{-}(x)=f(x-h)$ for forward and backward shifted
versions of any function $f$, and $\Delta f=h^{-1}(f_{+}-f)$ for the forward
difference operator. Ward's idea is to seek a discrete version of the
potential energy of the form
\beq
\label{6}
E_{P}=h\sum_{x}\left(\frac{1}{2}D^{2}+\frac{1}{8}F^{2}\right)
\eeq
where $D\ra\cd\phi/\cd x$ and $F\ra(1-\phi^{2})$ in the continuum limit
$h\ra 0$, and $D$ and $F$ have the product
\beq
\label{9}
DF=-\Delta\left(\frac{\phi^{3}}{3}-\phi\right).
\eeq
Such a lattice
potential energy clearly has the correct continuum limit. The crucial
point is that, given (\ref{9}), the discrete system has a Bogomol'nyi
argument exactly analogous to its continuum counterpart. Let
$\phi:h\Z\ra\R$ have kink boundary conditions. Then
\bea
0&\leq&\frac{h}{2}\sum_{x}\left(D-\frac{1}{2}F\right)^{2}=
E_{P}-\frac{h}{2}\sum_{x}DF \nonumber \\
&=&E_{P}+\frac{h}{2}\sum_{x}\Delta\left(\frac{1}{3}\phi^{3}-\phi\right)
=E_{P}-\frac{2}{3},
\eea
using the boundary conditions $\phi\ra\pm 1$ as $x\ra\pm\infty$. It follows
that
$E_{P}\geq \frac{2}{3}$ with equality if and only if $\phi$ satisfies the
lattice Bogomol'nyi equation,
\beq
\label{11}
D=\frac{1}{2}F.
\eeq
The most natural choice of $D$ and $F$ seems to be \cite{Spe}
\bea
D&=&\Delta\phi \nonumber \\
F&=&1-\frac{1}{3}(\phi_{+}^{2}+\phi_{+}\phi+\phi^{2}),
\eea
which clearly have the correct continuum limits, and are easily shown to
satisfy equation (\ref{9}). Note that the lattice Bogomol'nyi equation
(\ref{11}) is a first order difference equation, while the static field
equation,
\beq
\frac{\cd E_{P}}{\cd\phi}=0
\eeq
is a second order difference equation. We have shown that {\em if} a solution
of (\ref{11}) exists with kink boundary conditions then it is a global 
minimal of $E_{P}$ (within
the kink sector) with energy $\frac{2}{3}$, and hence a static kink
solution. The next section is devoted to proving the existence of such
solutions. 

We will refer to this lattice system as the
topological discrete $\phi^{4}$ system (TD$\phi^{4}$S) since it preserves
the topological lower bound on kink energy.  The conventional discrete
$\phi^{4}$ system also has a potential energy of the form (\ref{6}), but
while $D$ is the same (a standard forward difference), $F$ is chosen to be
the same as in the continuum model, $F=1-\phi^{2}$. In this case, the
Bogomo'nyi argument is lost. So the essential difference here is that we
have ``distributed'' the double well substrate potential over pairs of
nearest neighbour lattice sites. In both discrete systems, one completes the
dynamics by defining the obvious kinetic energy functional,
\beq
E_{K}=\frac{h}{2}\sum_{x}\dot{\phi}^{2}.
\eeq
Other interesting choices can be made \cite{Zak}, but this is the simplest.

\section{The kink moduli space}
\label{sec:kin}

We start with a definition: a two-sided sequence $\phi:h\Z\ra\R$ is a static
kink if it is a solution of the Bogomol'nyi equation (\ref{11}), is
monotonic and converges to $1$ at $\infty$ and $-1$ at $-\infty$. The space
of all static kinks is called the kink moduli space, denoted, for a particular
value of $h$, by $M_{h}$. Given that discretization has broken the continuous 
translation symmetry of the continuum model to symmetry under integer
translations by $h$, one would expect $M_{h}$ to be discrete. This is indeed
true for conventional discrete systems, where kinks must lie halfway
between or exactly on top of lattice sites, the difference in energy between
 these
two types of static solution being the origin of the PN barrier, which so
complicates the kink dynamics. However, the \T, remarkably, has a
continuous moduli space: static kinks may take any position relative to the 
lattice. All such kinks saturate the Bogomol'nyi bound on energy, so there
is no PN barrier resisting their propagation through the lattice. Specifically,
we will prove,
\begin{prop}
If $h\in(0,2]$ then for all $\phi_{0}\in(-1,1)$ there exists a unique
static kink with $\phi(0)=\phi_{0}$. 
\end{prop}
Note that a continuous translation orbit of kinks can equally well be 
parametrized by kink position $b\in\R$ (defined, for example by linear
interpolation) or by the value of the field at $x=0$, $\phi(0)\in(-1,1)$.
So proposition 1 could be rephrased: for all $h\in(0,2]$, $M_{h}=\R$. 
This is the important result for our purposes. It is also interesting to
consider what happens to the kink moduli space for $h>2$. In fact, except
for some special values of $h\in(2,3]$, $M_{h}=\emptyset$. However, for
an infinite discrete subset of $(2,3]$, $M_{h}=h\Z$. Precisely,
\begin{prop} For $h>3$ there exist no static kinks. There exists a strictly
decreasing sequence $h_{n}$ with $h_{0}=3$ and $\lim_{n\ra\infty}h_{n}=2$
such that if $h=h_{n}$ there exists exactly one static kink (modulo lattice
translations). This kink is symmetric about its central site(s) and has exactly
$n+1$ sites different from $\pm 1$.
\end{prop}
{\bf Proof of proposition 1:} Since the Bogomol'nyi equation,
\beq
\label{16.5}
\phi_{+}^{2}+\left(\frac{6}{h}+\phi\right)\phi_{+}+
\left(\phi^{2}-\frac{6}{h}\phi-3\right)=0
\eeq
is forward backward symmetric, that is, invariant under $(\phi_{+},\phi)\mapsto
(-\phi,-\phi_{+})$, it suffices to prove that for any $h\in(0,2]$ the right
hand sequence converges to $1$ for all $\phi(0)\in(-1,1)$. Note that given
$\phi\in(-1,1)$, equation (\ref{16.5}) does not uniquely determine $\phi_{+}$
since both roots are real. However, the lower root is always less than $\phi$,
and we seek monotonic solutions, so we can discard the lower root at each
iteration. It follows that if a static kink with $\phi(0)=\phi_{0}\in(-1,1)$
exists, it is unique.

It remains to prove convergence to 1 of the sequence 
\beq
\phi(nh)=f^{n}(\phi_{0})
\eeq
generated by iterating upper root of equation (\ref{16.5}), starting at any
value $\phi_{0}\in(-1,1)$ and assuming that $0<h\leq 2$. Here $f^{n}$ denotes
the $n$-th composition of $f$ (so $f^{0}={\rm Id}$), and
\beq
f(q)=\frac{1}{2}\left[-\left(\frac{6}{h}+q\right)+
\sqrt{\left(\frac{6}{h}+q\right)^{2}-4\left(q^{2}-\frac{6}{h}q-3\right)}\right]
\eeq
A straightforward calculation shows that on $[-1,1]$,
$f(q)\geq q$, the only fixed points being $\pm 1$.
Furthermore, $f(q)>1$ if and only if $q\in I_{h}$, this being the interval
$(\gamma-|\gamma-1|,\gamma+|\gamma-1|)$ where $\gamma=3h^{-1}-2^{-1}$. Since
$I_{h}\cap[-1,1]=\emptyset$ if $h\leq 2$, we find that $f$ is bounded above by
$1$ on $[-1,1]$. So the sequence $f^{n}(\phi_{0})$ is monotonic and bounded, 
and hence, by the monotone convergence theorem, convergent. Call its limit 
$L$. We have established that $f:[-1,1]\ra[-1,1]$, so the sequence 
$f^{n}(\phi_{0})$ lies in the interval $[-1,1]$. Either there exists 
$N\in\N$ such
that $f^{n}(\phi_{0})=L$ for all $n\geq N$ or $L$ is an accumulation point
of $[-1,1]$. In either case $L\in[-1,1]$ since $[-1,1]$ is closed. Consider
then the image of the sequence $f^{n}(\phi_{0})$ under $f$. Since $f$ is
continuous, $f(f^{n}(\phi_{0}))\ra f(L)$. But $f(f^{n}(\phi_{0}))$ is a 
subsequence of $f^{n}(\phi_{0})$, so converges to $L$. Hence $L$ is a fixed
point of $f$, and since $L$ cannot be $-1$, it follows that
$\phi(nh)\ra 1$.
 $\Box$

\vspace{1cm}
It is interesting to compare the situation with that of the topological
discrete sine-Gordon (TDSG) system \cite{Spe}. There, one is fortunate enough
to have an explicit formula for the static kink solution, obtained by
slightly modifying the continuum kink. One finds that $M_{h}=\R$ for all
$h<2$, but the ``proof'' of this fact rests on the observation that the
explicit formula has a continuous real-valued parameter in it. Unfortunately,
no continuum-inspired ansatz seems to work for the \T Bogomol'nyi equation,
so we are forced to give a less constructive proof. Note that in both systems
the kink moduli space degenerates at $h=2$. This may seem an artificial
coincidence which would be lost if different normalizations were used, but 
in fact the \T was normalized specifically so that the minimum phonon
frequency (the angular frequency of small amplitude travelling waves) is unity,
as in the TDSG system. One then finds that both systems have the {\em same}
phonon dispersion relation,
\beq
\omega^{2}=\frac{4+h^{2}}{4h^{2}}-\frac{4-h^{2}}{2h^{2}}\cos kh
\eeq
obtained by substituting a travelling wave ansatz $\cos(kx-\omega t)$
with frequency $\omega$ and wavenumber $k$  into the linearized equation
of motion (linearized about the vacuum $\phi=1$, not $\phi=0$). So in both
systems $M_{h}$ degenerates precisely where the phonon dispersion relation
collapses to a flat line. Since there is no obvious relation between the 
existence
of kinks, a nonlinear phenomenon, and the radiation of the system, a linear
phenomenon, this coincidence is rather strange. One {\em difference} between
the two systems is that, whereas there are no kinks for $h>2$ in the
TDSG system, there are kinks in the \T for certain discrete values of $h$
in $(2,3]$, as stated in proposition 2, to whose proof we now turn.

\vspace{1cm}
\noindent
{\bf Proof of proposition 2:} We adopt the definitions and notation used in
the proof of proposition 1. As stated above, $f(q)>1$ if $q\in I_{h}$, so
a static kink must have $\phi(0)\notin I_{h}$. Consider the backwards iteration
$\phi_{-}=\tilde{f}(\phi)$. By forward-backward symmetry, $\tilde{f}(q)<-1$ if
$q\in\tilde{I}_{h}=(-\gamma-|\gamma-1|,-\gamma+|\gamma-1|)$, where
$\gamma$ was defined above. So any static kink must also have $\phi(0)\notin
\tilde{I}_{h}$. If $h>3$ then $(-1,1)\subset I_{h}\cup\tilde{I}_{h}$. Hence
there exists no static kink for $h>3$.

Let $h\in(2,3]$ and $\phi(x)$ be a static kink. Since $\lim_{x\ra\pm\infty}
\phi=\pm 1$, either $\phi(\pm x)=\pm 1$ for all $x>Nh$ for some $N\in\Z^{+}$, 
or either $1$ or $-1$ is an accumulation point of the sequence. 
The latter cannot be the
case since $\phi$ is bounded away from $1$ and $-1$ by the intrusion of $I_{h}$
and $\tilde{I}_{h}$ into $[-1,1]$ respectively (see figure 1). 
So the static kink must be
constant away from its centre. This is possible if and only if there exists
$n\in\N$ such that
\beq
\label{18}
f^{n}(-\gamma+|\gamma-1|)=\gamma-|\gamma-1|
\eeq
where $f^{n}$ again denotes the $n$-th composition of $f$. 
For then if $\phi(0)=-\gamma+|\gamma-1|$, the backward
sequence jumps immediately to $-1$, which is a fixed point, and the forward
sequence leads, after $n$ iterations to $\gamma-|\gamma-1|$, which is
subsequently mapped to the fixed point $1$. Clearly such a kink is unique
modulo lattice translations, is symmetric and has 
$n+1$ sites different from $\pm 1$. For example, if $h=3$ then
$-\gamma+|\gamma-1|=0$ so there is a static kink with $\phi(0)=0$ and 
$\phi(x)=x/|x|$ elsewhere. The only other static kinks are obtained by
lattice translation of this one. 

So for each $n\in\N$ there is a $n+1$ site kink in the system with spacing
$h=h_{n}$ where $h_{n}$ is a solution of equation (\ref{18}); $h_{0}=3$ is a 
solution
with $n=0$. We seek to prove the existence of a decreasing sequence of such
solutions, $(h_{n})$, converging to $2$. Let $\chi=\gamma-|\gamma-1|$, so
$h\in[2,3]\Rightarrow\chi\in[0,1]$, and define $g(\chi)=-f(-\chi)$, a
graph of which is presented in figure 2. Then the condition (\ref{18}) for the 
existence of a static kink can be rewritten
\beq
\label{20}
g^{n}(\chi)=-\chi.
\eeq
Note that $g(0)=-1$, $g(1)=1$ and $g$ is continuous, so $g$ has a root,
call it $r_{1}$, between $0$ and $1$, and so there exists a solution $\chi_{1}$
of $g(\chi)=-\chi$ in $(0,r_{1})$. Now, if $g^{n}$ has a root $r_{n}\in(0,1)$,
then $g^{n+1}$ has a root $r_{n+1}\in(r_{n},1)$ since $g^{n+1}(r_{n})=-1$
and $g^{n+1}(1)=1$. Hence, by induction, for each $n\in\Z^{+}$, $g^{n}$ has
a root $r_{n}\in(r_{n-1},1)$, and since $g^{n}(r_{n-1})=-1$ there exists a
solution $\chi_{n}$ of equation (\ref{20}) with $\chi_{n}\in(r_{n-1},r_{n})$.
The sequence $(\chi_{n})$ is monotonic and bounded, hence convergent, and
$\lim_{n\ra\infty}\chi_{n}=\lim_{n\ra\infty}r_{n}$.

It remains to show that $\lim_{n\ra\infty}r_{n}=1$, for then the corresponding
sequence of $h$ values converges to $2$ as claimed. Note that $g$ is convex and
increasing on $[0,1]$, whence it follows that $\frac{1}{2}\leq r_{1}<1$
(consider the cord from $(0,-1)$ to $(1,1)$ in figure 2). 
Consider now the function
$g^{n}:[r_{n-1},1]\ra[-1,1]$ where $n\geq 2$. Defining $k:=g^{n-1}$, one
finds that
\beq
(g^{n})''(\chi)=g''(k(\chi))(k'(\chi))^{2}+g'(k(\chi))k''(\chi).
\eeq
The first term is nonnegative since $g$ is convex on $[0,1]$, and if
$k$ is convex on $[r_{n-2},1]$ then the second term is nonnegative
since $g'\circ k>0$ on $[r_{n-1},1]$ ($g$ is increasing on $[0,1]$). Hence,
by induction, $g^{n}:[r_{n-1},1]\ra[0,1]$ is convex, for each $n\in\Z^{+}$. 
It follows that $1-2^{-n}\leq r_{n}<1$ and hence that $\lim_{n\ra\infty}r_{n}
=1$. $\Box$

\vspace{1cm}
For $h\in(2,3]$ then, $M_{h}$ is discrete. One should not, however, interpret
this as the appearance of a PN potential in the system, since the discreteness
stems from our requiring that static kinks be monotonic (so that no site may
lie outside $[-1,1]$). Numerical solution of the lattice Bogoml'nyi equation
suggests that even if $h>2$ there {\em are} solutions with the correct 
boundary behaviour (and hence $E_{P}=\frac{2}{3}$) for any $\phi_{0}\in(-1,1)$,
but that they ``overshoot'' and approach the vacua from the wrong sides
(i.e.\ as $x\ra\infty$, $\phi(x)\ra 1$ from above, rather than below). We do
not regard these as being real kink solutions since they are not even
qualitatively close to continuum kinks sampled on the lattice.
The conclusion is, then, 
 that to simulate the continuum theory one should use a 
lattice with $h<2$.
Note that choosing, for example, $h=1.5$, the system is in an extremely
discrete regime since the kink structure is spread over very few lattice sites
(see figure 3).

\section{Numerical simulations}
\label{sec:num}

Having established that kinks of the \T experience no PN barrier, the question 
arises whether the resulting kink dynamics is
consequently simpler than that customarily observed in discrete systems. We 
address this question by performing numerical simulations of the lattice
equation of motion. This involves approximately  solving the following system 
of coupled nonlinear o.d.e's,
\bea
\ddot{\phi}&=&\frac{1}{h^{2}}(\phi_{+}-2\phi+\phi_{-}) \nonumber \\
        & &\qquad+\frac{1}{12}\left[(2\phi+\phi_{-})\left(1-\frac{1}{3}
(\phi^{2}+\phi\phi_{-}+\phi_{-}^{2})\right)+\right. \nonumber \\
& &\left.\qquad\quad\,\,\,\,\,\,\,
(2\phi+\phi_{+})\left(1-\frac{1}{3}(\phi^{2}+\phi\phi_{+}+
\phi_{+}^{2})\right)\right]
\eea
on a large (but finite) lattice, with a Lorentz boosted {\em continuum}
kink solution as initial data. (Although it would be straightforward to
generate a genuine discrete static kink for the initial profile, there is no
obvious way to boost it, without an explicit formula. In any case, the point
of the endeavour is to simulate continuum kink dynamics.) Simulations were
performed with a variety of lattice spacings and initial kink velocities.
The solutions were generated using a simple fourth order Runge-Kutta scheme. 

In every case it was found that the kink moves
 freely through the lattice, without
being pinned (see figure 4), exciting small amplitude travelling waves which
propagate backwards in its wake. To reduce reflexion of this radiation from
the fixed left hand boundary during long simulations, the first few lattice
sites were damped. Figure 5 presents plots of velocity against time for
several long duration simulations. (The thickness of the curve is due to
velocity oscillations as the kink passes from cell to cell. This dynamical
wobble is not caused by any PN type potential, but can be understood in terms
of geodesic motion on the moduli space $M_{h}$ with respect to a periodic
metric \cite{Spe}.) After an initial drop in velocity as the kink relaxes to a profile more suited to the discrete system, subsequent kink deceleration is
very modest, even at high speed ($v=0.6$) and on coarse lattices.

This behaviour should be contrasted with that of the standard discrete
system, whose equation of motion,
\beq
\ddot{\phi}=\frac{1}{h^{2}}(\phi_{+}-2\phi+\phi_{-})
+\frac{1}{2}\phi(1-\phi^{2})
\eeq
was solved using the same algorithm and initial data, for purposes of 
comparison (also figure 5). These results are similar to those obtained by
Combs and Yip \cite{Com} in their work on the conventional discrete $\phi^{4}$
system. One sees that at every value of $h$, kink dynamics
in the \T is far less dissipative, the difference being more pronounced
for larger $h$. Indeed, a kink set off with a speed $v=0.6$ on a lattice with
$h=1.8$ quickly becomes trapped in the conventional system (figure 6), but
remains free (apparently) indefinitely in the \T.

\section{Concluding remarks}
\label{sec:con}

In this paper we have shown that by choosing a discretization which preserves 
the Bogomol'nyi bound on kink energy, one can find a discrete $\phi^{4}$ system
which has no Peierls-Nabarro potential. The resulting kink dynamics is much 
simpler than one expects of a discrete system, even on coarse lattices. The
kinks are never pinned, and suffer only modest radiative deceleration. So the
\T is an efficient and natural choice of discretization for numerical study
of the continuum $\phi^{4}$ model.

In higher dimensional field theories with topological solitons (the $O(3)$
sigma, Skyrme and abelian Higgs models for example) the reduction in
computational cost would be quite significant if similarly natural
discretizations could be found. Here, however, there is an extra concern:
solitons on high dimensional lattices tend to be unstable because they can
unwind and ``fall through'' a single plaquette. One way to preserve the
stability of the continuum model is again to preserve the Bogomol'nyi
bound on soliton energy, an approach pursued by Ward \cite{War} for two
planar field theories. In this case it seems impossible to eliminate the PN
barrier entirely, although one can attempt to reduce it as much as possible.
It would be interesting to see to what extent these discrete systems have
continuum-like soliton dynamics, as has been found here for $\phi^{4}$ kinks,
and elsewhere \cite{Spe,Zak} for sine-Gordon kinks.

\vspace{2cm}
\noindent
\large
{\bf Figure captions}
\normalsize
\vspace{1cm}
\newline Figure 1: The shaded region, excluding boundaries, represents
$\phi$ values which get mapped outside the interval $[-1,1]$ under either
forward or backward iteration of the Bogomol'nyi equation. The intersection
of any vertical line with this region is $I_{h}\cup\tilde{I}_{h}$. 
\vspace{0.5cm}
\newline Figure 2: The function $g(\chi)$ of the proof of proposition 2.
\vspace{0.5cm}
\newline Figure 3: A static kink of the model with $h=1.5$. Circles represent
the kink profile itself, while the solid line represents its energy 
distribution, which should be regarded as residing in the links between
lattice sites. Note the extreme discreteness of this structure: over 99\%
of the kink's energy is carried by the 4 central links.
\vspace{0.5cm}
\newline Figure 4: Propagation of a kink through the lattice of spacing 
$h=1$, with an initial speed $v=0.2$. Note the apparent lack of radiation in
the kink's wake.
\vspace{0.5cm}
\newline Figure 5: Comparison of the motion of fast kinks (initial velocity
$v=0.6$) in the \T with that
in the conventional discrete system, for lattices of spacing $h=1.0$, $h=1.4$
and $h=1.8$. In each case velocity is plotted against time for both systems on
the same graph,
the upper curve showing the \T data and the lower curve showing the 
conventional system data. The difference is particularly striking for $h=1.8$,
since the kink quickly becomes pinned in the conventional system, while it 
remains free indefinitely in the \T. 
\vspace{0.5cm}
\newline Figure 6: Kink pinning in the conventional system with $h=1.8$.
Set off at a speed of $0.6$, the kink manages to travel through only 8
lattice cells before being trapped by the PN barrier. This should be compared
with the dynamics of a \T kink on the $h=1.8$ lattice, which remains free
indefinitely and after $3000$ time units has decelerated by less than 17\%.

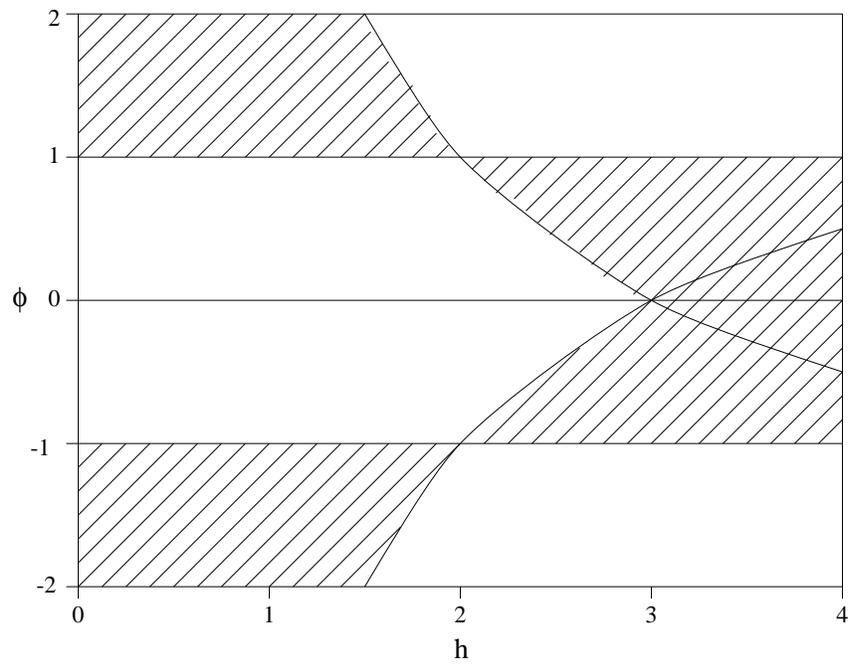
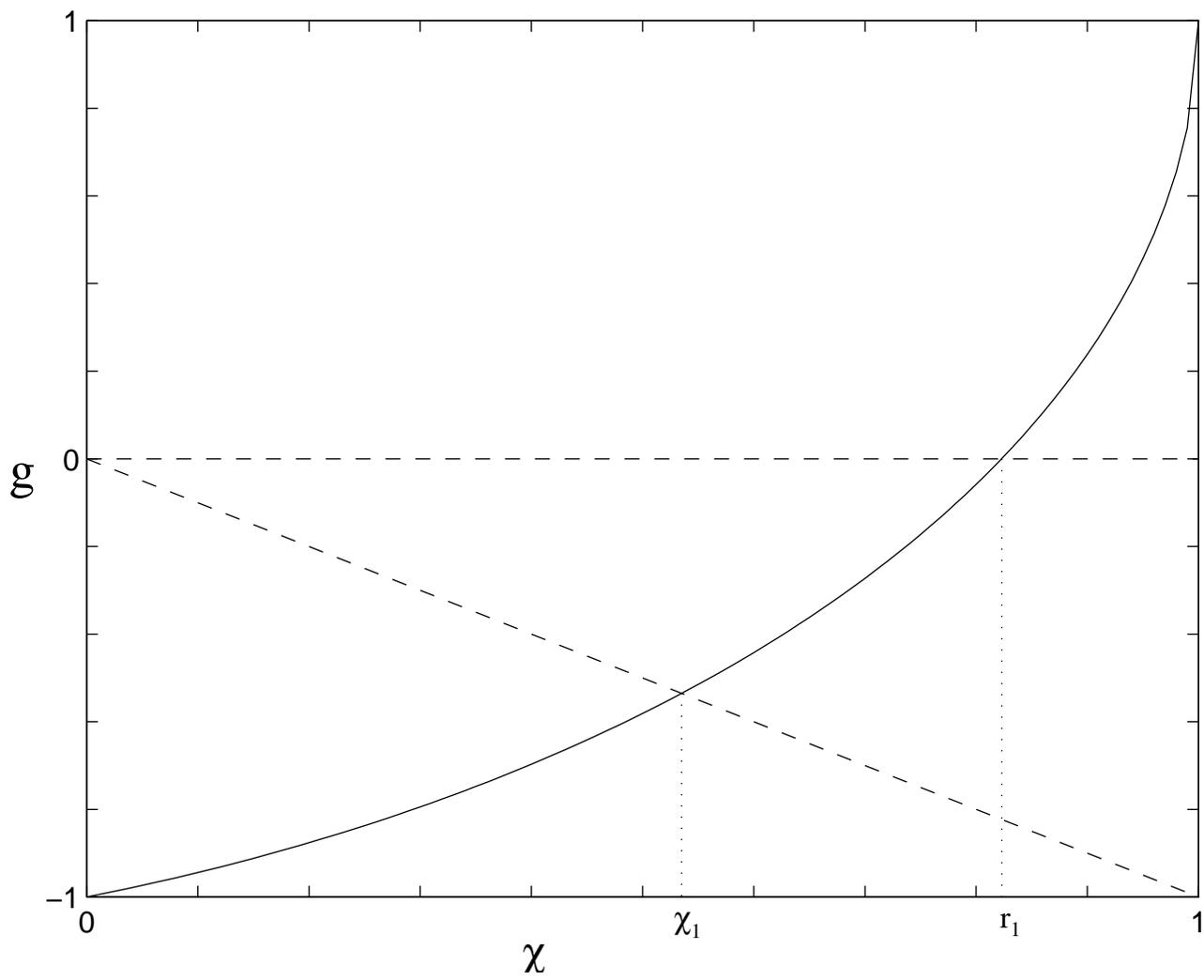
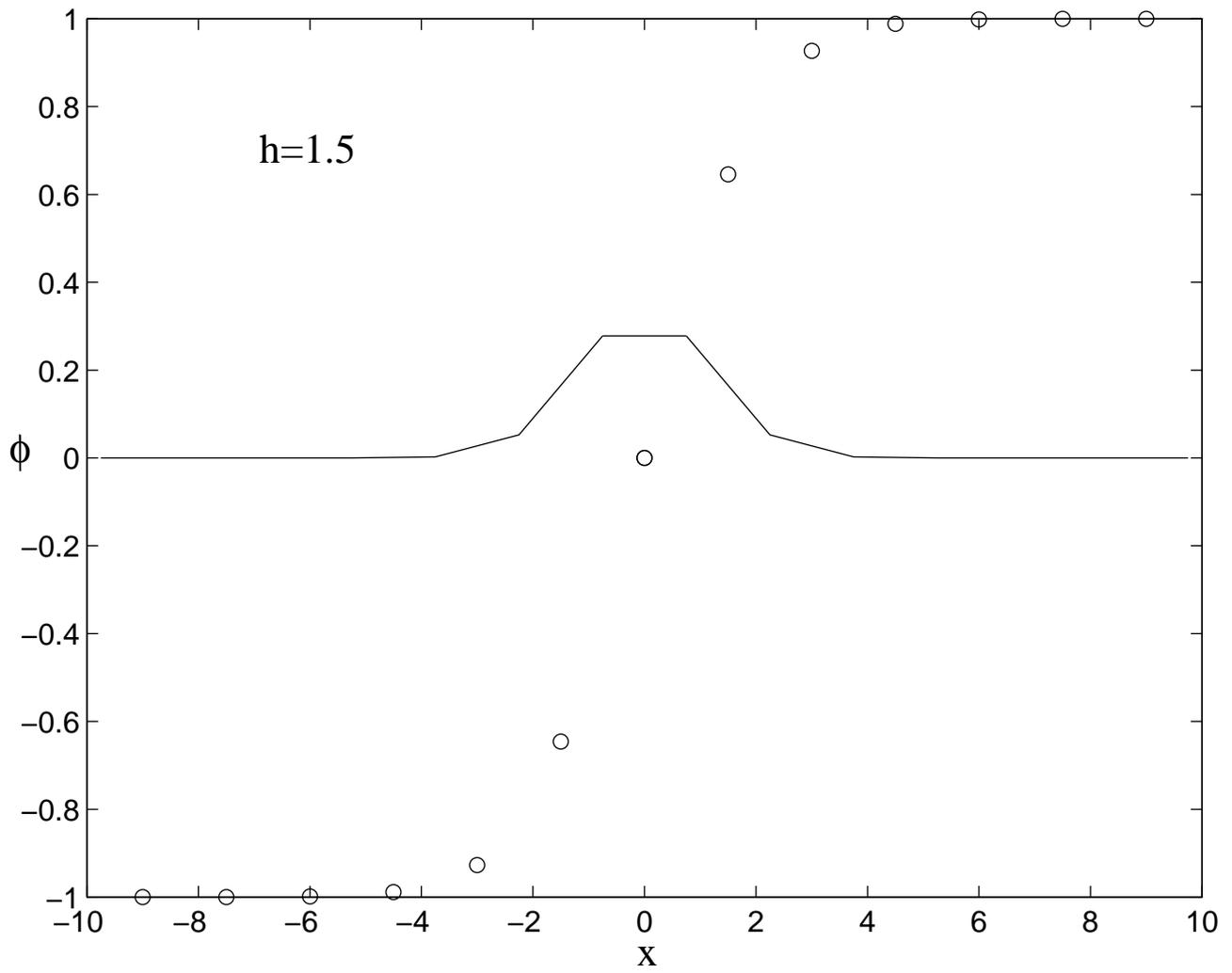
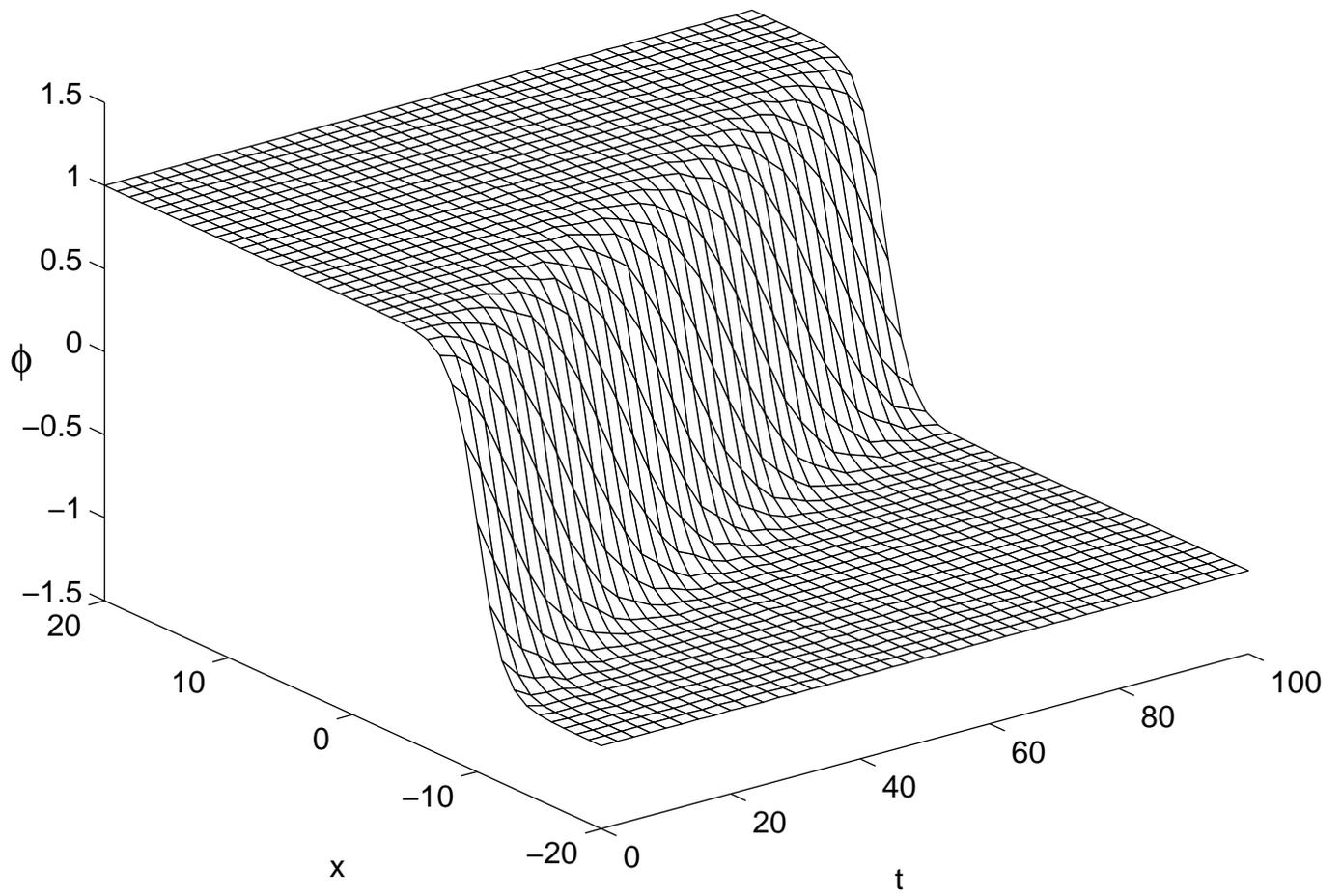
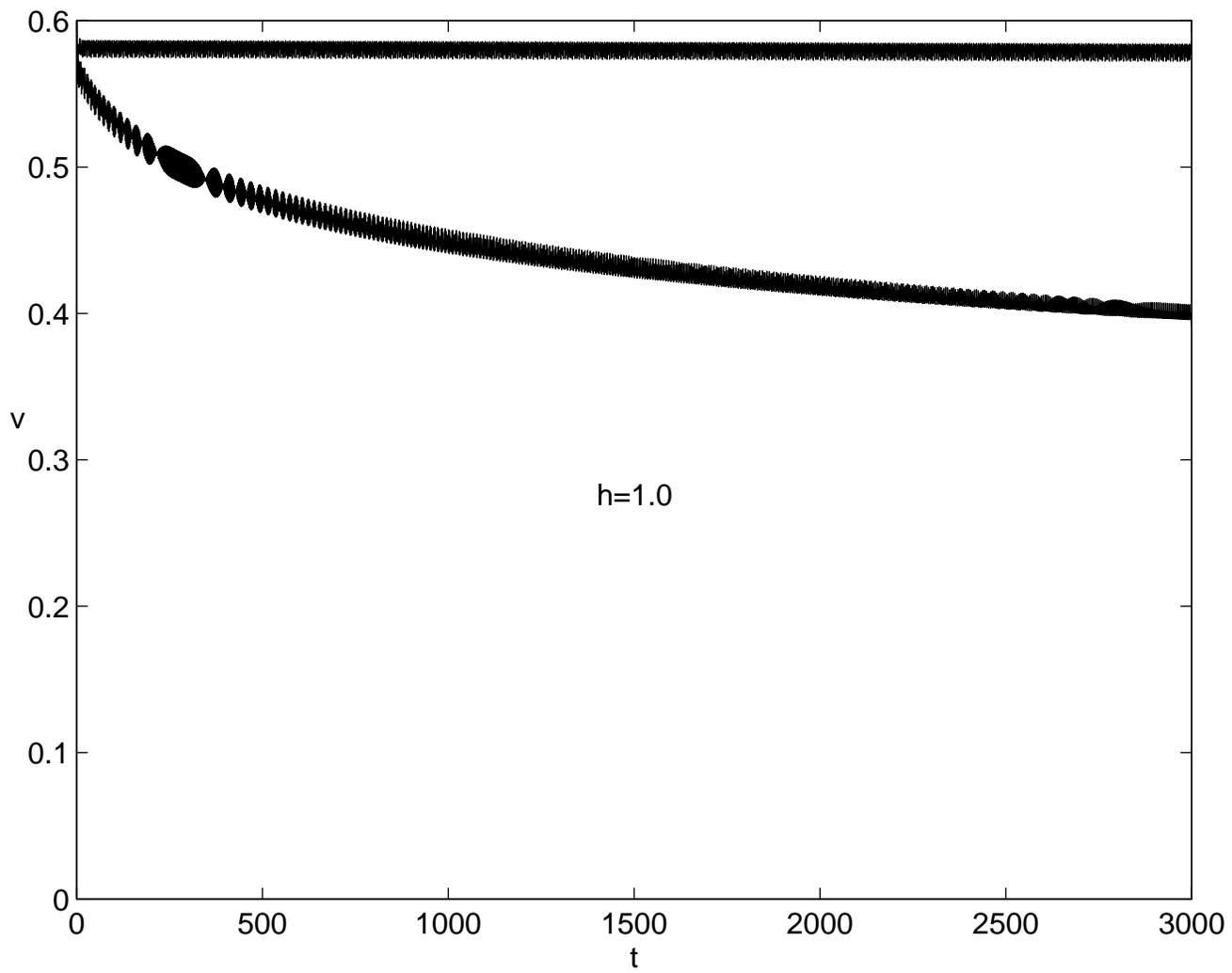
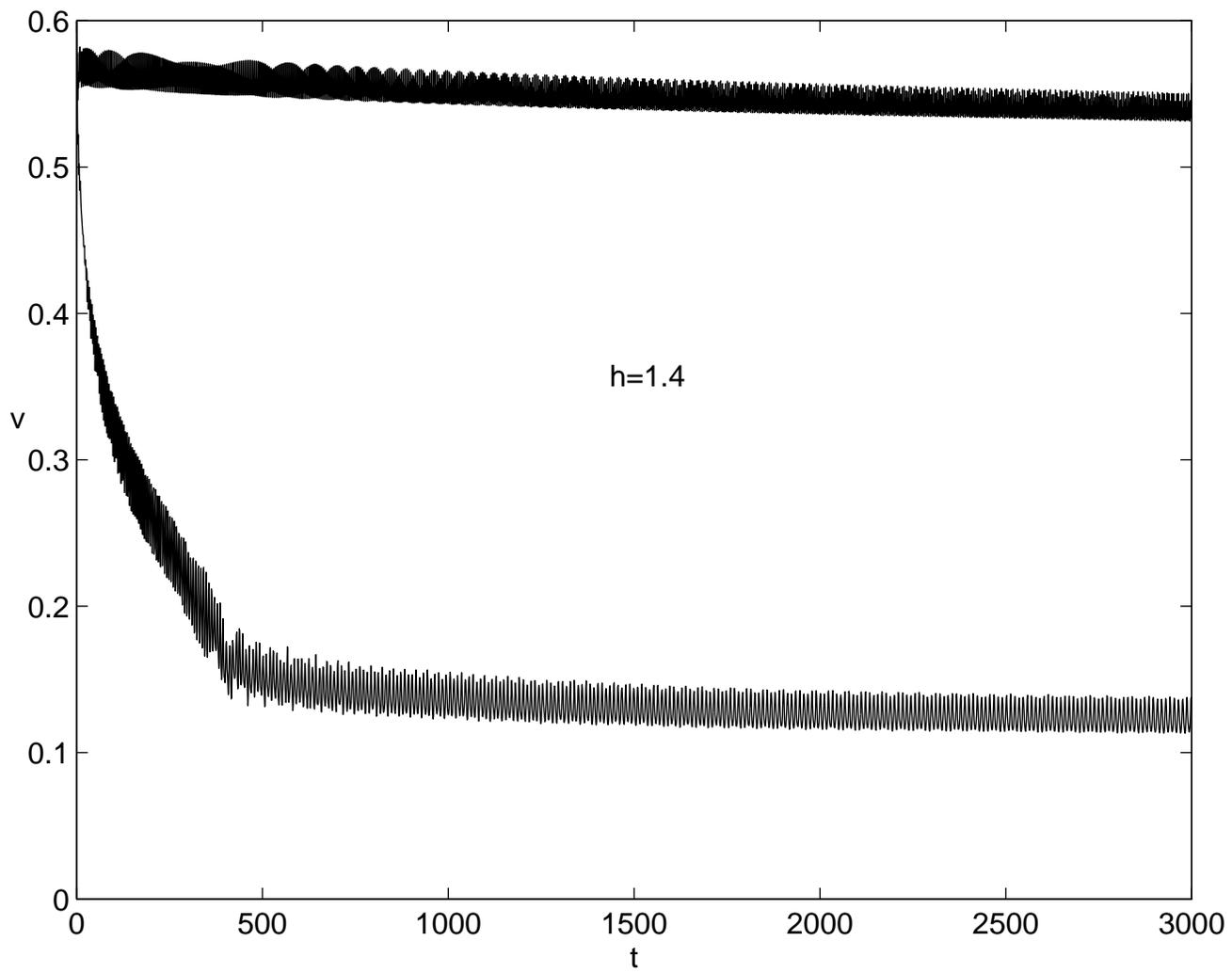
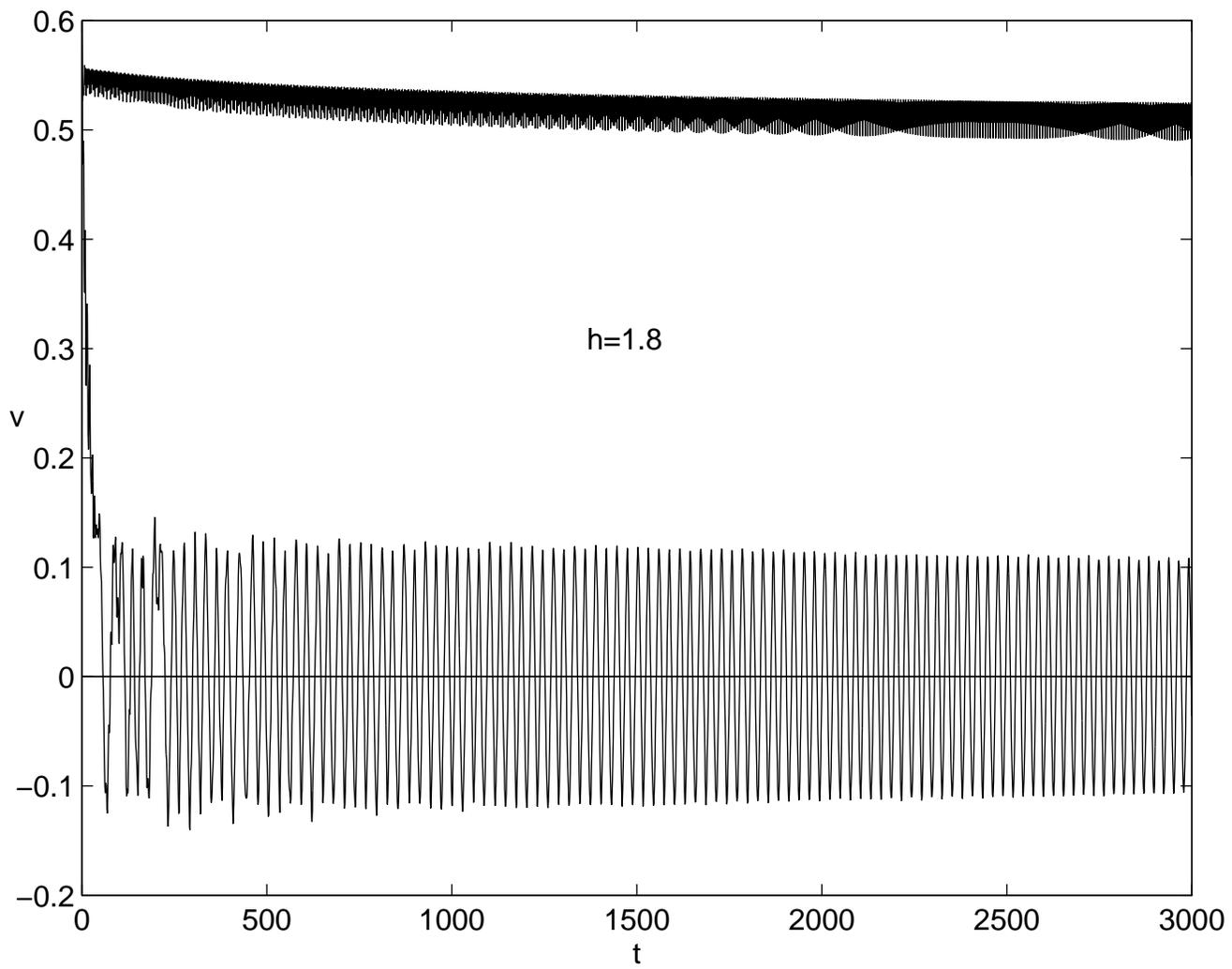
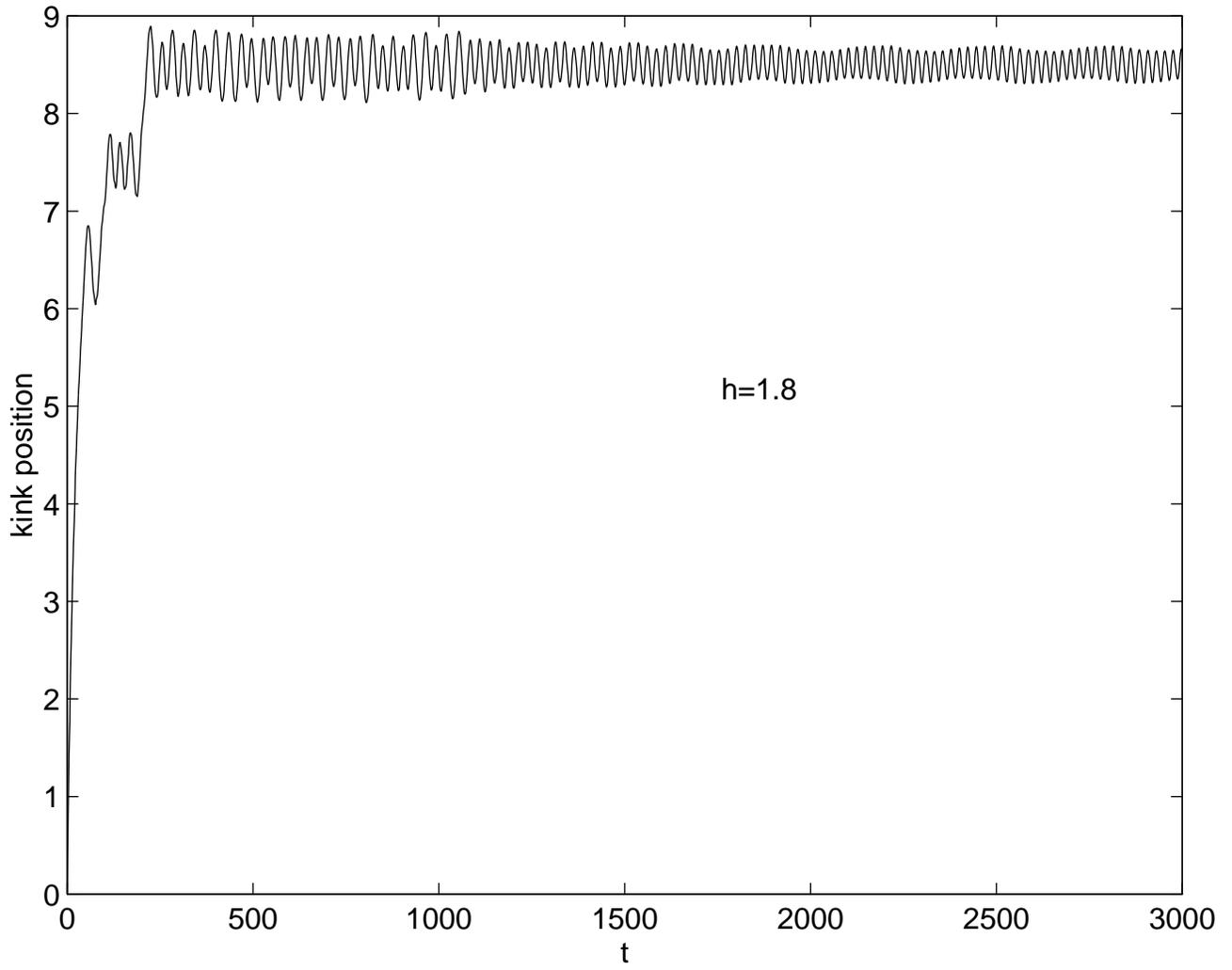
\end{document}